\documentclass[aps,prl,reprint,groupedaddress,longbibliography]{revtex4-1}
\usepackage{graphicx,amsmath,amssymb,braket,siunitx}
\DeclareSIUnit\gauss{G}
\DeclareSIUnit\sig{\mbox{$\sigma$}}

\begin{document}

\title{Deep laser cooling and efficient magnetic compression of molecules}
\author{L. Caldwell}
\author{J. A. Devlin}\altaffiliation{Present address: CERN, 1211 Geneva, Switzerland}
\author{H. J. Williams}
\author{N. J. Fitch}
\author{E. A. Hinds}
\author{B. E. Sauer}
\author{M. R. Tarbutt}
\email[]{m.tarbutt@imperial.ac.uk}
\affiliation{Centre for Cold Matter, Blackett Laboratory, Imperial College London, Prince Consort Road, London SW7 2AZ UK
}

\begin{abstract}
We introduce a scheme for deep laser cooling of molecules based on robust dark states at zero velocity. By simulating this scheme, we show it to be a widely applicable method that can reach the recoil limit or below. We demonstrate and characterise the method experimentally, reaching a temperature of \SI{5.4+-0.7}{\micro\kelvin}. We solve a general problem of measuring low temperatures for large clouds by rotating the phase-space distribution and then directly imaging the complete velocity distribution. Using the same phase-space rotation method, we rapidly compress the cloud. Applying the cooling method a second time, we compress both the position and velocity distributions.
\end{abstract}

\maketitle
There has been rapid progress in laser cooling of molecules in recent years. Several species have been cooled~\cite{Shuman2010,Hummon2013,Zhelyazkova2014,Prehn2016,Kozyryev2017,Lim2018}, and 3D magneto-optical traps (MOTs) have been demonstrated for a few~\cite{Barry2014,Truppe2017,Williams2017,Anderegg2017,Collopy2018}. Sub-Doppler cooling~\cite{Truppe2017, Cheuk2018}, and internal state control~\cite{Williams2018,Blackmore2018} have been developed, and the molecules stored in magnetic and optical traps~\cite{Williams2018,McCarron2018,Anderegg2018}. These laser-cooled molecules can be used to test fundamental physics~\cite{Safronova2018,DeMille2017}, simulate many-body quantum systems~\cite{Micheli2006}, process quantum information~\cite{DeMille2002,Yelin2006,Andre2006}, and study quantum chemistry~\cite{Krems2008}. All these applications require, or would benefit from, lower temperatures and higher densities. Here, we address both requirements, demonstrating a method that cools CaF molecules to \SI{5}{\micro\kelvin}, and a method that efficiently compresses the cloud. We show that these two techniques can work together to increase the density {\it and} reduce the temperature. Finally, we present a technique that directly measures the complete velocity distribution of the ultracold sample. This thermometry is superior to the standard ballistic expansion method when the distribution is non-thermal, as happens, for example, with deep cooling schemes~\cite{Lawall1995}, velocity selection schemes~\cite{Fox2005}, narrow-line MOTs~\cite{Grunert2002} and ultracold plasmas~\cite{Killian2007}. It is also superior whenever distributions are cold but large, as happens in the present work and other applications including atom interferometry~\cite{Kovachy2015} and narrow-line MOTs~\cite{Grunert2002}.

\begin{figure*}[tb]
    \includegraphics[width=\textwidth]{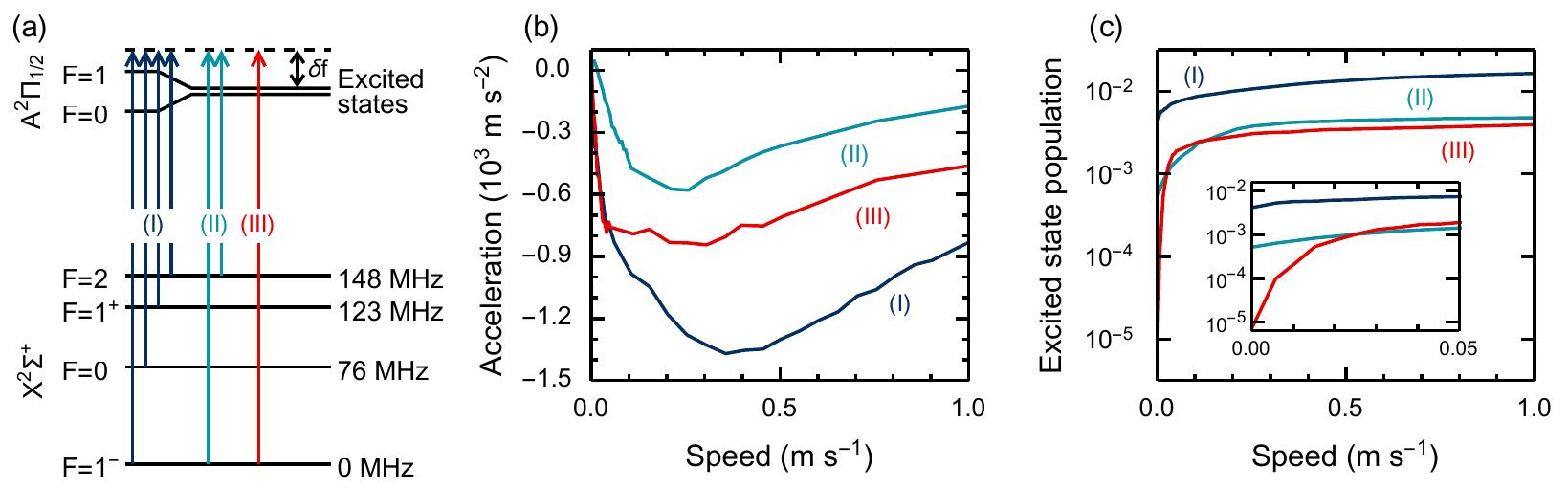}
    \caption{(a) Hyperfine components of the laser cooling transition in CaF, with three cooling schemes shown: (I) multi-frequency; (II) $\Lambda$-enhanced; (III) single-frequency. (b) Steady-state acceleration versus speed and (c) excited-state fraction versus speed, obtained from optical Bloch equation simulations of each scheme. Simulation parameters are: (I) total intensity $I=117$~mW/cm$^2$, detuning $\delta f = 20$~MHz; (II) $I=50$~mW/cm$^2$, $\delta f=30$~MHz; (III) $I=340$~mW/cm$^2$, $\delta f=8.3$~MHz. Parameters for (I) and (II) are close to those which give the lowest measured temperatures~\cite{Truppe2017, Cheuk2018}.}
    \label{fig:simulations}
\end{figure*}

First, we explain our new laser-cooling method. Figure \ref{fig:simulations}(a) shows the hyperfine structure of the $A^2\Pi_{1/2}\leftarrow X^2\Sigma^+$ laser cooling transition in CaF at \SI{606.3}{\nano\metre}. Further details of the level structure are given in Supplemental Materials (SM)~\cite{Supp}. To make a MOT~\cite{Truppe2017,Williams2017,Anderegg2017}, radio-frequency sidebands are applied to the cooling light to address each of the ground state hyperfine components. The hyperfine structure of the excited state is not resolved. When the magnetic field is off and all sidebands are blue-detuned, sub-Doppler cooling is effective and the molecules cool to \SI{55}{\micro\kelvin}~\cite{Truppe2017, McCarron2018, Anderegg2018}. This multi-frequency molasses is shown as scheme (I) in Fig.~\ref{fig:simulations}(a). The molasses temperature is limited by the momentum diffusion arising from photon scattering.  A good way to reduce this is to engineer a robust dark state\textemdash{}an eigenstate of the Hamiltonian that is not coupled by the light to any excited state\textemdash{}for molecules at zero velocity. This velocity selective coherent population trapping turns off the heating for the slowest particles, and has been used to cool atoms below the recoil limit~\cite{Aspect1988,Lawall1995}. Sometimes, the dark state is produced using a two-photon resonance between two hyperfine states. This method, often called $\Lambda$-enhanced gray molasses~\cite{Grier2013}, was recently used by \citet{Cheuk2018} to cool CaF to \SI{5}{\micro\kelvin}. As shown by scheme (II) of Fig.~\ref{fig:simulations}(a), they turned off two sidebands and tuned the remaining two into resonance with the Raman transition between $F=1^-$ and $F=2$, to engineer dark states that are superpositions of these levels. Nevertheless, a significant scattering rate remained, corresponding to an excited-state fraction of $1.3(2)\times 10^{-3}$. This is because the dark states are de-stabilized by off-resonant excitation (the frequency and polarization component targeting one hyperfine state may also excite the other), and by the $F=0$ level in the excited state which couples to $F=1^-$ but not to $F=2$. This suggests that even lower temperatures might be reached by finding more robust velocity-selective dark states, while retaining a strong cooling force. Our approach to achieve this is to use a single frequency component, blue detuned from all hyperfine levels, as illustrated by scheme (III) in Fig.~\ref{fig:simulations}(a). For stationary molecules, and for any polarization, there are two dark states that are superpositions of $F=2$ Zeeman sub-levels. One is an eigenstate of the full Hamiltonian, including the kinetic energy operator. Moving molecules spend some of their time in bright states, where the average light shift can be large because the light has high intensity and is not too far detuned from $F=1^{-}$. Consequently, there can be a strong cooling force, but little scattering, for molecules at low speeds. These are the requirements for efficient 3D cooling to the recoil limit and below~\cite{Lawall1995, Papoff1992}.

To explore these ideas, we simulate all three schemes illustrated in Fig.~\ref{fig:simulations}(a). We use 3D optical Bloch equation simulations that include all relevant molecular levels and laser frequency components, and all six beams of the molasses~\cite{Devlin2016,Devlin2018}. The motion of the molecules is treated classically. Results of these simulations are shown in Fig.~\ref{fig:simulations}(b,c). Figure \ref{fig:simulations}(b) shows that while scheme I gives the largest force over the widest range of speeds, scheme III provides just as high a damping constant at low speed. Figure \ref{fig:simulations}(c) shows that, at all speeds, schemes (II) and (III) have lower excited-state population than scheme (I), and that in scheme (III) this drops to very low values at the lowest speeds, because the population is pumped into stable dark states at zero velocity~\footnote{In the simulations, the cooling time is \SI{0.8}{\milli\second}. In schemes (I) and (II) the populations have reached their steady-state values in this time, whereas in (III) the excited-state fraction continues to fall for longer cooling times.}. This opens the possibility of cooling below the recoil limit, which would not be possible with the other schemes where a substantial scattering rate remains even at zero speed. Further discussion of the dark states involved in (II) and (III) is given in the SM~\cite{Supp}. Using the data in Figs.~\ref{fig:simulations}(b,c), and the Fokker-Planck-Kramers equation~\cite{Devlin2018}, we estimate a lower temperature limit, $T_{\rm low}$, for each scheme. For (I), we predict $T_{\rm low} = \SI{11+-1}{\micro\kelvin}$, about 4 times lower than measured. The discrepancy arises because our method neglects heating due to fluctuations of the dipole force~\cite{Devlin2018}. For (II), we predict $T_{\rm low} = \SI{5.4+-0.8}{\micro\kelvin}$, consistent with measurements~\cite{Cheuk2018}. For (III), the predicted temperature is below the recoil limit of \SI{0.44}{\micro\kelvin}. In this regime a full quantum mechanical treatment of the motion is needed.

Our experiments begin with a cloud of $2\times10^4$ CaF molecules cooled to $\sim$ \SI{55}{\micro\kelvin} by scheme (I)~\cite{Williams2017,Truppe2017}. To implement scheme (III), we switch off the cooling light, turn off the modulators that add sidebands to the laser, step the laser frequency so that the detuning from $F=1^-$ is $\delta  f$, and then after a settling period, turn the light back on with intensity $I$. The period between one molasses and the other is \SI{700}{\micro\second}. The repump lasers remain on at full intensity throughout. After holding the molecules in the single-frequency molasses for time $t_{\rm sfm}$, we measure the temperature, $T$.

\begin{figure*}[tb]
    \includegraphics[width=\textwidth]{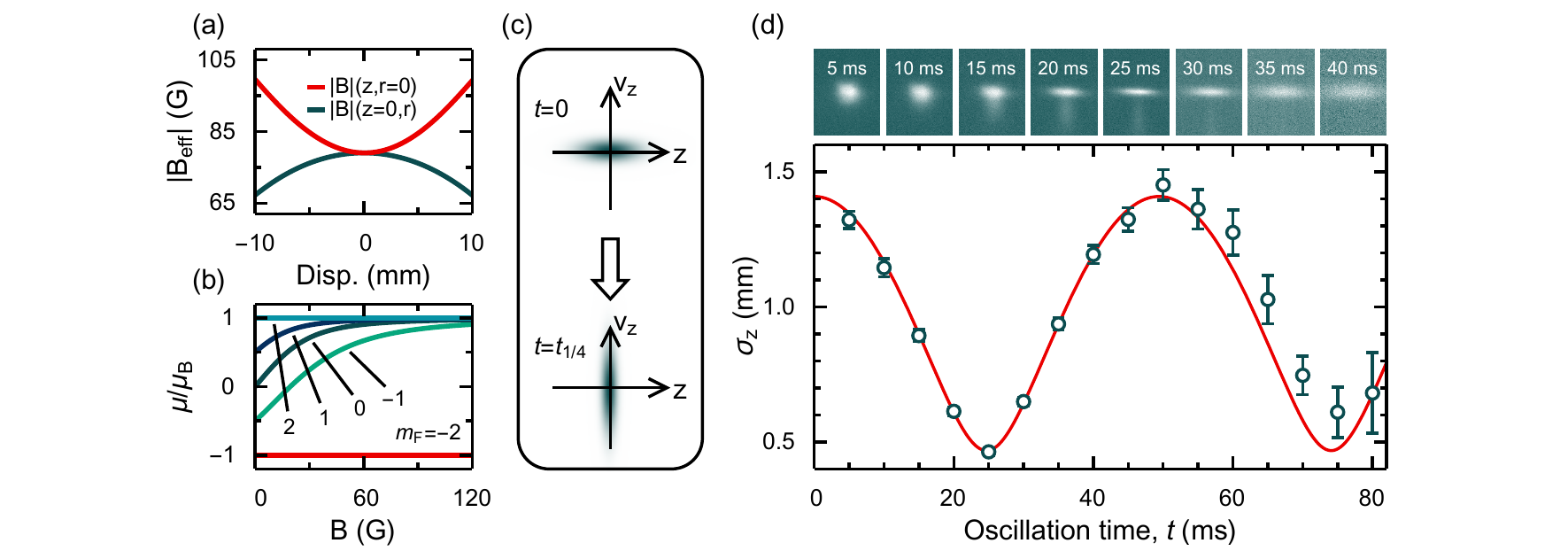}
    \caption{Phase space rotation in a 1D magnetic trap. (a) Magnitude of effective magnetic field, $|B_{\rm eff}|=|B| + m g z/\mu_{\rm B}$, versus axial and radial displacements. Weak-field seeking molecules are trapped axially. (b) Magnetic moments, $\mu$, of Zeeman sub-levels in the $X^2\Sigma^+(N=1, F=2)$ state, versus $B$. At large $B$, $\mu\approx \mu_{\rm B}$ for $m_F \in \{-1, 0, 1, 2\}$. (c) Phase space evolution in harmonic trap. After a quarter period, the position distribution is proportional to the initial velocity distribution. (d) Top: fluorescence images of molecules at \SI{5}{\milli\second} intervals, averaged over 50 shots; $z$ is vertical, and the field of view is \SI{13.5}{\milli\meter} wide. Bottom: rms width in $z$ direction versus time in trap, $\sigma_{z}(t)$. Error bars include systematic uncertainties. Line is a fit to Eq.~(\ref{eqn:width}).}
    \label{fig:phase-space-rotation}
\end{figure*}

We first measured $T$ using the standard ballistic expansion method. Using optimised molasses parameters (see later), we measure axial and radial temperatures of \SI{8+-1}{\micro\kelvin} and \SI{6.7+-0.6}{\micro\kelvin} by this method, where we have given statistical uncertainties only. At these low temperatures, the cloud expands by less than \SI{1}{\milli\metre} in the time taken to leave the field of view, so the size is always dominated by the initial size ($\sigma_0 >\SI{1}{\milli\metre}$), and the velocity distribution is never clearly revealed. In this situation, small deviations from a Gaussian spatial distribution, and imaging aberrations near the edges of the field of view, can lead to systematic shifts that dominate the temperature measurements. Indeed, in the axial direction ($z$), our data do not fit perfectly to the ballistic expansion model (see SM~\cite{Supp}). 
The expansion time could be increased by magnetically levitating the cloud, but this introduces large systematic uncertainties due to the distribution of magnetic moments, even at large bias fields. 

Because of these difficulties, we use the method illustrated in Fig.~\ref{fig:phase-space-rotation} to measure the axial temperature. After cooling, we turn off the light and apply a magnetic field, $B$, which has an offset $B_0 \approx \SI{80}{\gauss}$, a gradient along $z$ that cancels gravity for molecules with magnetic moment $\mu=\mu_{\rm B}$, and a curvature along $z$ that traps these molecules with an angular oscillation frequency $\omega$. The method for making this field is described in the SM~\cite{Supp}. Our simulations predict that, after cooling, the molecules are uniformly distributed amongst the $m_F$ levels of $F=2$, with no population in other states. Fig.~\ref{fig:phase-space-rotation}(b) shows $\mu$ vs $B$ for these states, and the corresponding Zeeman shifts are shown in the SM~\cite{Supp}. At $B\approx B_0$, molecules in $m_F=-2$  have $\mu=-\mu_{\rm B}$ and are ejected, while all others have $\mu \approx \mu_{\rm B}$ and are confined. As illustrated in Fig.~\ref{fig:phase-space-rotation}(c), their phase-space distribution rotates in the harmonic trap, so by imaging the cloud after quarter of a period, we measure the initial velocity distribution. This is true for any initial distribution in phase space. When the initial velocity distribution is thermal with temperature $T$, and the initial position distribution is Gaussian with rms width $\sigma_0$, the position distribution remains Gaussian at all times $t$, with an rms width of
\begin{equation}
	\sigma(t) = \sqrt{\sigma_0^2\cos^2{\omega t} + \frac{k_{\rm B} T}{m \omega^2}\sin^2{\omega t}},\label{eqn:width}
\end{equation}
where $m$ is the mass. To measure this distribution, the magnetic field is turned off, the cooling light (with sidebands) immediately turned back on at the detuning used for the MOT, and the fluorescence imaged onto a CCD camera for \SI{750}{\micro\second}.

\begin{figure*}[t]
\includegraphics[width=\textwidth]{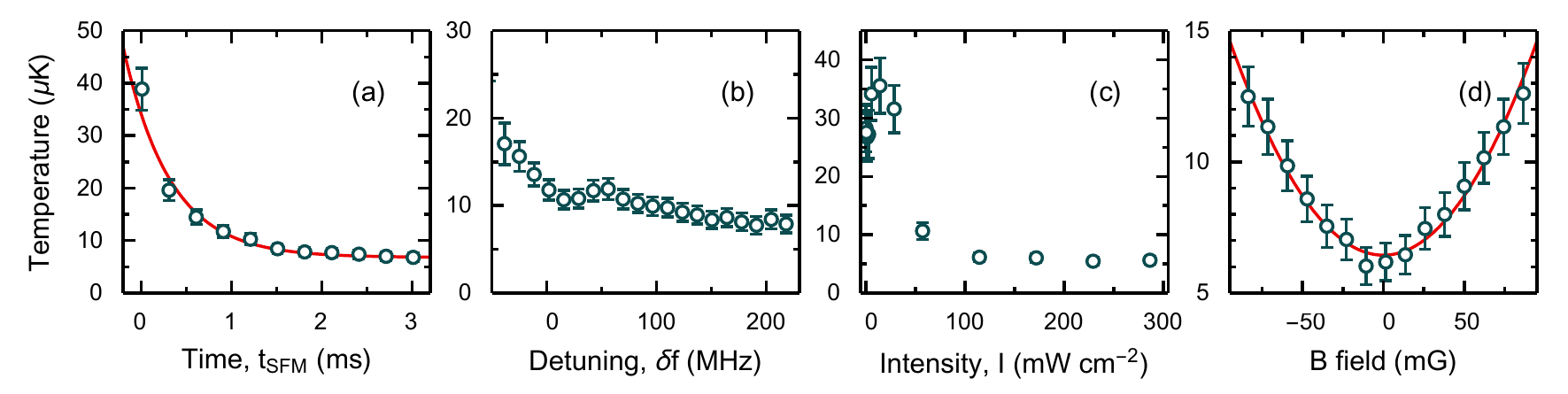}
\caption{Temperature versus single-frequency molasses parameters. (a) $T$ versus $t_{\rm sfm}$, when $I= \SI{287}{\milli\watt\per\centi\metre\squared}$ and $\delta f= \SI{164}{\mega\hertz}$. Line: fit to $T=T_0 + (T_{\rm i} - T_{0})e^{-t_{\rm sfm}/\tau}$, giving $\tau=\SI{0.52+-0.06}{\milli\second}$. (b) $T$ versus $\delta f$ when $t_{\rm sfm}=\SI{5}{\milli\second}$ and $I=\SI{287}{\milli\watt\per\centi\metre\squared}$. (c) $T$ versus $I$ when $t_{\rm sfm}=\SI{5}{\milli\second}$ and $\delta f= \SI{164}{\mega\hertz}$. (d) $T$ vs one component of $B$, after optimisation of the other two components; $t_{\rm sfm}=\SI{5}{\milli\second}$, $I=\SI{287}{\milli\watt\per\centi\metre\squared}$ and $\delta f= \SI{164}{\mega\hertz}$. Line: fit to $T=T_0+ \alpha B^{2}$, giving $\alpha = \SI{9+-1e-4}{\micro\kelvin\per\milli\gauss\squared}$. Points give average and standard error of 50 experimental runs. Error bars are dominated by systematic uncertainties.}
\label{fig:characterisation}
\end{figure*} 

The upper row of Fig.~\ref{fig:phase-space-rotation}(d) shows images at various times $t$. As expected, $\sim 80\%$ of the molecules are trapped axially and slowly stretched radially. The remaining molecules are squeezed radially and accelerated downwards.  We determine the rms width in $z$ of the trapped cloud, $\sigma_z$, by integrating the image along $r$ over a central region of width $w_{\rm cut}=\SI{4.43}{\milli\metre}$, then fitting to the resulting distribution. Using this central region reduces the effects of imaging aberrations near the edges of the field of view. For $t\leq \SI{30}{\milli\second}$, when the ejected molecules are still visible, we fit to the sum of two Gaussians with the position and width of each as free parameters. For later times a single Gaussian is sufficient. The lower panel of Fig.~\ref{fig:phase-space-rotation}(d) shows $\sigma_z$ versus $t$. The line is a fit to Eq.~(\ref{eqn:width}), giving $\omega/(2\pi) = \SI{10.12(2)}{\hertz}$, within 10\% of the value expected from a simple model of the coils, and $T=\SI{5.6+-0.6}{\micro\kelvin}$. This model fits well, apart from at late times where the measured size is inflated, mainly by the effect of defocussing as the cloud expands radially out of the imaging plane. This effect is reproduced by numerical modelling, and has a negligible effect on our determinations of $\omega$ and $T$. The temperature is related to the minimum width, $\sigma_{z,{\rm min}}$, obtained at time $t_{1/4}=\pi/(2\omega)$, by $k_{\rm B} T=m \omega^2 \sigma_{z,{\rm min}}^2 $. All subsequent temperature data are obtained from images taken at $t_{1/4}$ and integrated over $w_{\rm cut}/2$. The statistical uncertainty in a single measurement made this way is $\sim\SI{0.1}{\micro\kelvin}$, far smaller than when measured by ballistic expansion. Systematic shifts and uncertainties are discussed in the SM~\cite{Supp}, are accounted for in all data presented, and result in a correction of $\SI{-1.4(7)}{\micro\kelvin}$ for the coldest clouds.

Figure~\ref{fig:characterisation}(a) shows $T$ versus $t_{\rm sfm}$ along with a fit to an exponential decay giving a $1/e$ time constant of \SI{0.52+-0.06}{\milli\second}. A similar thermalisation time constant of \SI{0.41+-0.07}{\milli\second} is predicted by the simulations. Figure \ref{fig:characterisation}(b) shows how $T$ depends on $\delta f$. The temperature decreases as $\delta f$ is tuned from negative to positive, reaches its lowest values for $\delta f>\SI{180}{\mega\hertz}$, and is insensitive to $\delta f$ in this region, making the cooling robust. Simulations show the same dependence of $T$ on $\delta f$, including the mysterious bump near \SI{50}{\mega\hertz}, but at lower temperatures throughout. Figure~\ref{fig:characterisation}(c) shows that $T$ is insensitive to $I$ at high intensity, but increases at lower intensities which we attribute to a longer damping time at low $I$. Figure~\ref{fig:characterisation}(d) shows that $T$ varies quadratically with background magnetic field, with curvature $\SI{9+-1e-4}{\micro\kelvin\per\milli\gauss\squared}$. Simulations also show a quadratic dependence, but with a curvature 4 times higher. The lowest temperature is obtained when all three field components are zero (measured within \SI{10}{\milli\gauss}). After optimising all parameters, we measure $T=$\SI{5.4+-0.7}{\micro\kelvin}. This is consistent (within $2\sigma$) with the value measured by ballistic expansion, but more reliable for the reasons discussed above and in the SM~\cite{Supp}. Our simulations suggest that considerably lower temperatures are feasible, so we speculate that the temperature reached here may be limited by a time-varying magnetic field or laser polarisation, which could be improved. We observe no loss of molecules in the cooling step, other than the 20\% lost when measuring $T$, and the data fit well to a single thermal distribution, so we conclude that all the molecules are cooled to the same low temperature.

As seen in Fig.~\ref{fig:phase-space-rotation}(d), the cloud is compressed by a factor 3 at time $t_{1/4}$. It is common to compress magnetically-trapped atoms~\cite{Ketterle1999}, and recently molecules~\cite{McCarron2018}, by adiabatically increasing the trap frequency, $\omega$, or field gradient, $A$. Using this method, $\sigma$ scales only as $\omega^{-1/2}$ in a harmonic trap or as $A^{-1/3}$ in a quadrupole trap~\cite{Ketterle1999}. Thus, even modest reductions in cloud size require large field gradients, which must be maintained for long times to be adiabatic. Our rapid compression method is more effective because $\sigma$ scales as $\omega^{-1}$. The compression heats the cloud, but it can be re-cooled by applying the molasses a second time. This sequence increases the phase-space density provided (i) the cooling is fast enough that re-expansion during the second cooling phase is small, (ii) the velocities after compression are within the capture velocity of the molasses, and (iii) the magnetic field used for compression can be turned off rapidly enough that the molasses is effective. The first two conditions are easily satisfied, but the third is difficult. The $\sim\SI{80}{\gauss}$ field must be rapidly reduced below $\SI{50}{\milli\gauss}$, but eddy currents can produce large, slowly-decaying fields. We switch the coils off in a way that minimises these effects, as detailed in the SM~\cite{Supp}, then re-apply the single-frequency molasses for \SI{5}{\milli\second}, and finally measure $T$. Since the cloud is now much smaller, ballistic expansion measurements are reliable. Using this method, we measure $\sigma_0 = \SI{0.73+-0.01}{\milli\meter}$ and $T= \SI{14+-1}{\micro\kelvin}$, limited by the residual magnetic field, which could be reduced further. A longer wait between turning off the trap and re-applying the cooling reduces $T$ at the expense of $\sigma_0$. The complete cycle of cooling, compressing, and re-cooling reduces the axial size by a factor 2, reduces the temperature by a factor 4, and retains 80\% of the molecules. A greater $\omega$ will give stronger compression, and the compression improves if $T$ is lowered further since $\sigma \propto T^{1/2}$.

In summary, we have demonstrated a simple, robust method that cools molecules to \SI{5}{\micro\kelvin}. The method works by reducing the scattering rate to low values for the slowest molecules. Simulations suggest that temperatures down to the recoil limit of \SI{0.44}{\micro\kelvin}, or even lower, should be possible, and we are studying how to achieve that. This cooling should work for all molecular species laser cooled so far, and the principles for engineering robust velocity-selective dark states apply to new species too. We have developed a technique for directly measuring the velocity distribution of ultracold clouds that works for all phase-space distributions, and is superior to ballistic expansion for large clouds. Finally, we have shown that these ultracold clouds can be compressed using a conservative potential, and re-cooled after compression. For smaller clouds, the compression could be applied using an optical trap~\cite{Anderegg2018}. For example, a \SI{50}{\watt} laser at \SI{1}{\micro\meter} with a waist of \SI{300}{\micro\meter} would, we estimate, compress the cloud to $\sim\SI{90}{\micro\meter}$. The compression demonstrated in 1D can be extended by making a 3D harmonic potential, e.g. using a Ioffe-Pritchard trap. If the trap is isotropic, which is possible in this geometry~\cite{Bergeman1987}, the cloud can be compressed in all directions. Large increases in the 3D phase space density can be achieved by these methods. This is important for applications in quantum simulation and information processing, studies of ultracold collisions, and cooling molecules towards quantum degeneracy. 

Underlying data may be accessed from Zenodo \footnote{10.5281/zenodo.2564130} and
used under the Creative Commons CCZero license.

We are grateful to Simon Cornish for valuable discussions that led to some of the methods used in this work. We thank J. Dyne and V. Gerulis for expert technical assistance and A. Guo for work on characterising the imaging system. This work was supported by EPSRC under grants EP/M027716/1 and EP/P01058X/1.

\bibliography{references}

\onecolumngrid
\clearpage
\begin{center}
\textbf{\large Deep laser cooling and efficient magnetic compression of molecules: Supplemental Material}
\end{center}
\twocolumngrid
\setcounter{equation}{0}
\setcounter{figure}{0}
\setcounter{table}{0}
\makeatletter
\renewcommand{\theequation}{S\arabic{equation}}
\renewcommand{\thefigure}{S\arabic{figure}}
\renewcommand{\bibnumfmt}[1]{[S#1]}

\section{Energy levels relevant for laser cooling of CaF}

\begin{figure}
\includegraphics[width=0.4\columnwidth]{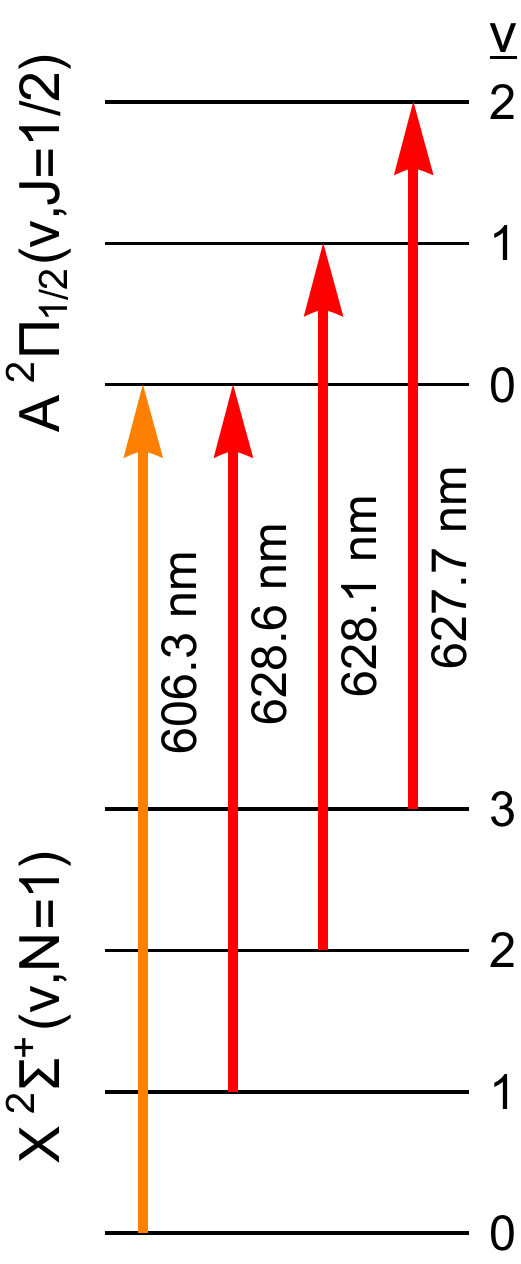}
\caption{Transitions involved in the laser cooling of CaF.\label{fig:vibLevels}}
\end{figure}

We use the standard notation where $\vec{L}$ is the orbital angular momentum operator, $\vec{S}$ is the electronic spin operator, $\vec{R}$ is the rotational angular momentum operator, $\vec{I}$ is the fluorine nuclear spin operator, $\vec{N} = \vec{L} + \vec{R}$, $\vec{J} = \vec{N} + \vec{S}$, $\vec{F} = \vec{J} + \vec{I}$, $m_{F}$ is the eigenvalue of $F_{z}$, $v$ is the vibrational quantum number and $p$ is the parity. Figure \ref{fig:vibLevels} shows the levels and transitions involved in the laser cooling scheme. The main cooling light drives the transition from $X ^{2}\Sigma^{+}(v=0, N=1)$ to $A ^{2}\Pi_{1/2}(v=0, J=1/2, p=+1)$. The transition has a wavelength of $\lambda = 606.3$~nm, and a natural width of $\Gamma = 2\pi\times 8.3$~MHz. If treated as a two-level system, the saturation intensity is $I_{\rm s} = \pi h c \Gamma/(3 \lambda^3) = 4.9$~mW/cm$^2$. Since it is not a two-level system, the actual intensity needed for the scattering rate to reach half its maximum value is higher, about 50~mW/cm$^{2}$~\cite{Williams2017}. The excited state cannot decay by an electric dipole transition to any other rotational states of $X ^{2}\Sigma^{+}$, because of the parity and angular momentum selection rules. It can, however, decay to other vibrational states, $v$, though with rapidly diminishing probability as $v$ increases. We use three vibrational repump lasers to address the transitions from $v=1,2$ and 3, as illustrated in Fig.~\ref{fig:vibLevels}.

\begin{figure}
\includegraphics[width=\columnwidth]{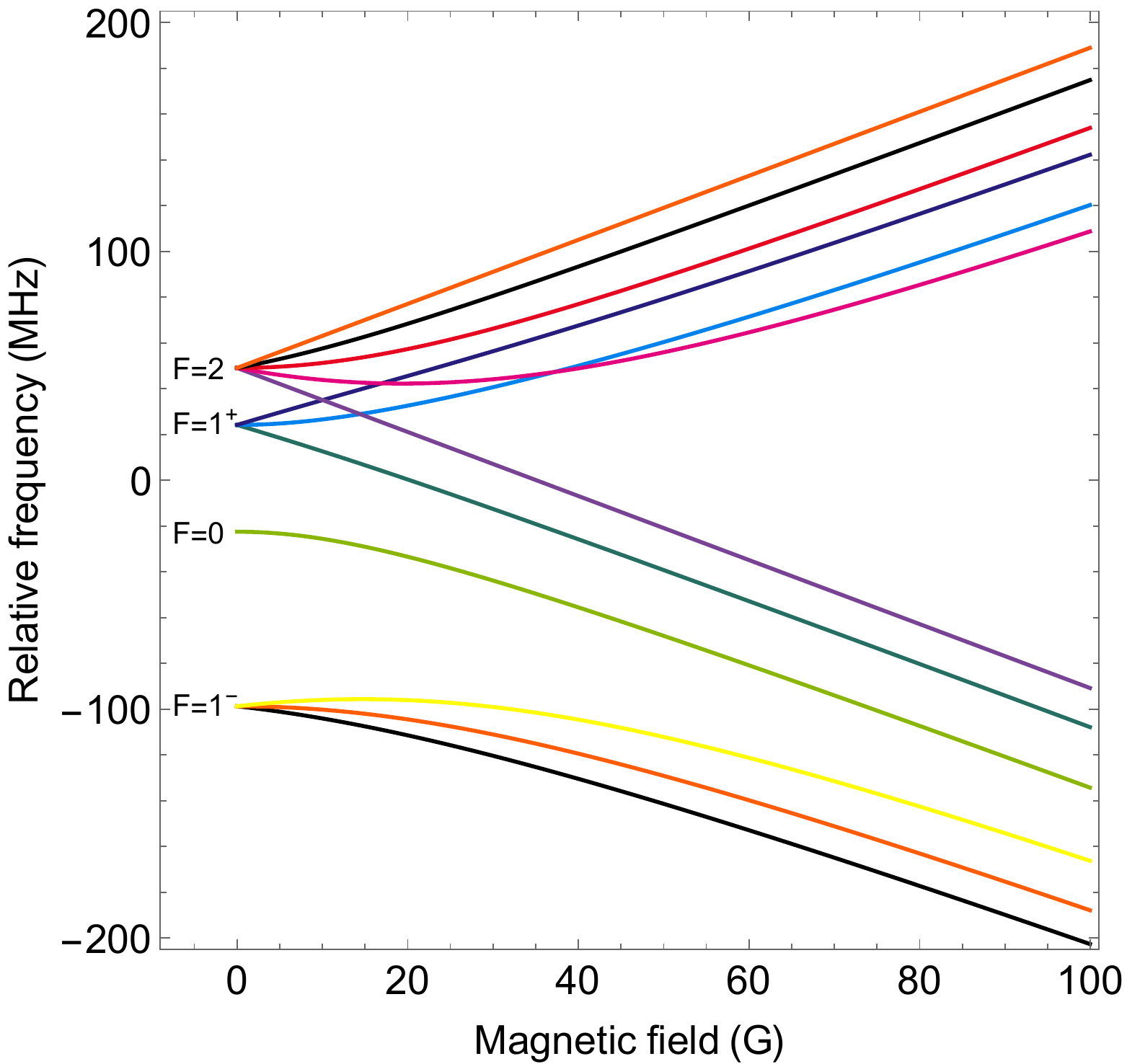}
\caption{Zeeman shifts of the four hyperfine components of the $X ^{2}\Sigma^{+}(v=0, N=1)$ ground state. \label{fig:zeemanLevels}}
\end{figure}

The ground state of the laser cooling transition is split by the spin-rotation interaction into two levels, $J=1/2$ and $J=3/2$. Each is further split into two by the hyperfine interaction to give a total of four levels $F=0,1,1,2$, as illustrated in Fig.~\ref{fig:simulations}(a). We refer to these as the four hyperfine components, and we distinguish the two $F=1$ levels using the notation $F=1^-$ and $F=1^+$, where the latter has higher energy. Figure \ref{fig:zeemanLevels} shows the Zeeman splittings of the four hyperfine components of $N=1$. At the offset field used in the experiment, $B_0 \approx 80$~G, the levels separate into two groups that have $m_S = \pm 1/2$. The molecules that have $m_S=1/2$ are confined by the axial harmonic trap, and all have magnetic moments close to $\mu_{\rm B}$. The excited state of the laser cooling transition is split by the hyperfine interaction into two levels, $F=0$ and $F=1$, which are not resolved in the experiment.

\section{Dark states}
\begin{figure}
\includegraphics[width=\columnwidth]{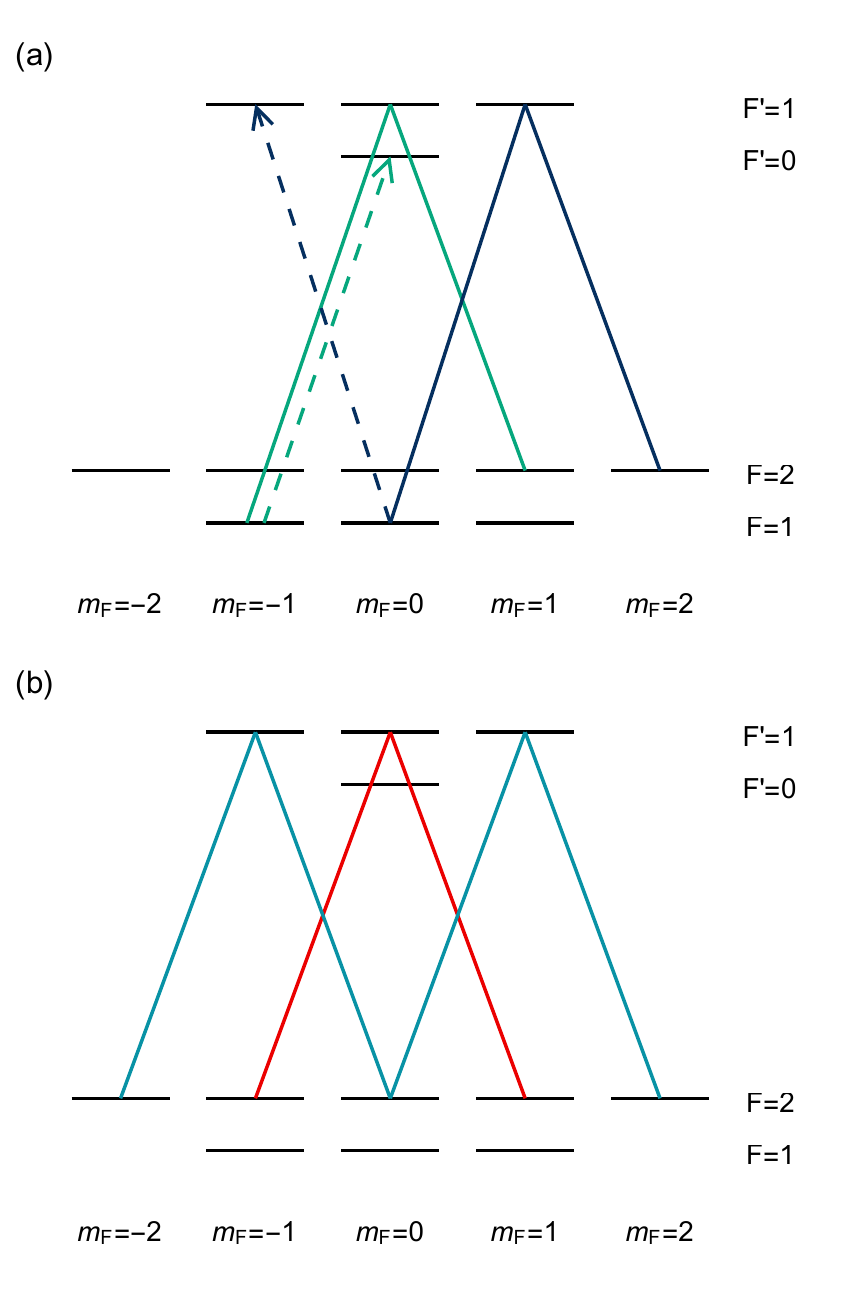}
\caption{(a) Quasi-dark states in the $\Lambda$-enhanced cooling scheme. One frequency component drives $\sigma^+$ transitions from $F=1$ while the other drives $\sigma^-$ transitions from $F=2$. The dark state shown in green is de-stabilized by excitation to the $F'=0$ excited state (dotted green arrow). The dark state shown in blue is de-stabilized by off-resonant-excitation (dashed blue arrow). (b) Dark states in a single-frequency $\sigma^+ \sigma^-$ molasses.\label{fig:darkStates}}
\end{figure}

Here, we consider in more detail the dark states that are formed in the $\Lambda$-enhanced gray molasses and the single-frequency molasses. We consider a ground state with $F=1$ and $F=2$ hyperfine levels, and an excited state with $F'=0$ and $F'=1$ hyperfine levels. The ground states are labelled $\ket{F,m_F}$ and the reduced matrix element for the transition between $F$ and $F'$ is $d_{F,F'}$. Figure \ref{fig:darkStates}(a) illustrates two potential dark states in the $\Lambda$-enhanced gray molasses. Here, we consider the prototypical case where one frequency component drives $\sigma^+$ transitions from $F=1$, and the other drives $\sigma^-$ transitions from $F=2$. There are two quasi dark states that are coherent superpositions of $F=1$ and $F=2$. They are $\ket{\psi_1} = \ket{2,2}-\sqrt{2}a\ket{1,0}$ (shown in blue) and $\ket{\psi_2}=\ket{2,1}-a\ket{1,-1}$ (shown in green), where $a=\sqrt{\frac{3}{5}}\frac{d_{2,1}}{d_{1,1}}$ and we have omitted the normalisation. Neither of them is completely dark. $\ket{\psi_1}$ is not completely dark because of off-resonant excitation of the $\ket{1,0}$ state, as illustrated by the dashed blue line. $\ket{\psi_2}$ is not completely dark because the $\ket{1,-1}$ state can be excited to $F'=0$, as illustrated by the dashed green line. Consequently, the scattering rate does not drop to zero in this scheme, even at zero velocity. These considerations extend to other laser polarizations and into 3D. There are no stable dark states, except in the case where the light field is unable to drive either $\sigma^{+}$ or $\sigma^{-}$ transitions. 

Figure \ref{fig:darkStates}(b) shows stable dark states in the single-frequency molasses. They are $\ket{2,1}-\ket{2,-1}$ (shown in red) and $\ket{2,2}-\sqrt{6}\ket{2,0}+\ket{2,-2}$ (shown in light blue). If the atomic motion is treated classically, both are true dark states. In a fully quantum-mechanical picture, where states are $\ket{F,m_F,p}$, and $p$ is the atomic momentum, the former remains a dark state, though the latter does not. Specifically, for counter-propagating $\sigma^{+}\sigma^{-}$ beams in one-dimension, where the $\sigma^{+}$ beam is parallel to $p$, the state $\ket{2,1,\hbar k}-\ket{2,-1,-\hbar k}$ is an eigenstate of the Hamiltonian, including the kinetic energy operator, and does not couple to the excited states. This then is a velocity-selective coherent population trapping state that can facilitate cooling to below the recoil limit~\cite{Papoff1992}. These considerations generalize to the three dimensional arrangement of counter-propagating circularly polarised laser beams used in the single frequency molasses~\cite{Lawall1995}. Note that in the presence of both $F'=1$ and $F'=0$ excited states, there are no equivalent dark states within the $F=1$ ground state. Note also that the dark states present in the single frequency molasses are absent in the $\Lambda$-enhanced molasses due to off-resonant excitation (the two different frequency components do not satisfy the Raman resonance condition for these states).

\section{Magnetic field control}

The MOT coils of inner diameter $\sim\SI{30}{\milli\metre}$ and separation $\sim\SI{38}{\milli\metre}$ are mounted inside the vacuum chamber. When connected with currents flowing in opposite directions, they create the quadrupole magnetic field needed for the MOT. We use the same coils to make the axial harmonic trap used to compress the cloud and measure the temperature. Using an H-bridge circuit we switch the direction of current in one of the coils. In this configuration, the coils produce a large offset magnetic field in the $z$-direction with a quadratic dependence on position along the z axis as shown in Fig.~\ref{fig:phase-space-rotation}(a). By additionally shunting some of the current around the top coil, we add a small field gradient which cancels the force due to gravity, so that the minimum of the total potential experienced by the molecules coincides with the centre of the two coils and the position of the MOT. The measured oscillation frequency in the axial harmonic trap is within 10\% of the value expected from a simple model of the coils. The magnetic field turns on to within 5\% of its final value in less than \SI{1.4}{\milli\second}.

\begin{figure}
\includegraphics{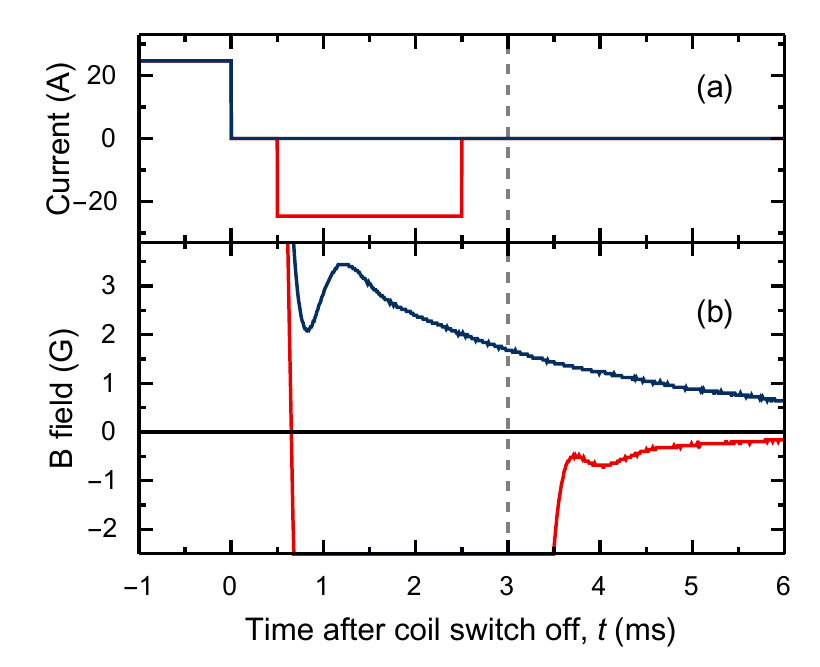}
\caption{Sequence for reducing eddy-currents when switching off harmonic trap. (a) Commanded current profile as a function of time. (b) Magnetic field measured on a Hall probe placed at the position of the molecules as a function of time. Blue lines: Standard switching. Red lines: extra step to reduce eddy-currents. Where not visible, red line is below blue. To make the harmonic trap, each coil carries \SI{25}{\ampere} of current in the same direction. This current is turned off at $t=0$ in the figure. To re-cool the molecules after compression, the single-frequency molasses light is turned on at $t=\SI{3}{\milli\second}$ (indicated by the dashed grey line). The aim is to make the magnetic field as small as possible during the \SI{5}{\milli\second}-long molasses period. The brief reversal of the current direction (red line) helps to achieve this. \label{fig:eddy-currents}}
\end{figure}

Eddy currents make it difficult to turn off the magnetic field rapidly. This problem is much worse when the currents in the two coils flow in the same direction compared to when they are opposite. To reduce the eddy currents, we apply the sequence illustrated in Fig.~\ref{fig:eddy-currents}(a). The current in the bottom coil is switched from \SI{25}{\ampere} to zero at $t=0$, then to \SI{-25}{\ampere} between $t=\SI{0.5}{\milli\second}$ and $t=\SI{2.5}{\milli\second}$, then back to zero. The current in the top coil flows in the same direction as the bottom coil and is always proportional to it, though smaller because of the shunt. The cooling light is turned on between $t=3$ and \SI{8}{\milli\second} for the second period of single-frequency molasses. Figure \ref{fig:eddy-currents}(b) shows the resulting magnetic field at the position of the molecules. We see that the magnetic field during the molasses period is about 3 times smaller when the reverse-current pulse is used.

\section{Systematic errors in temperature measurements}

\begin{figure}[tb]
\includegraphics{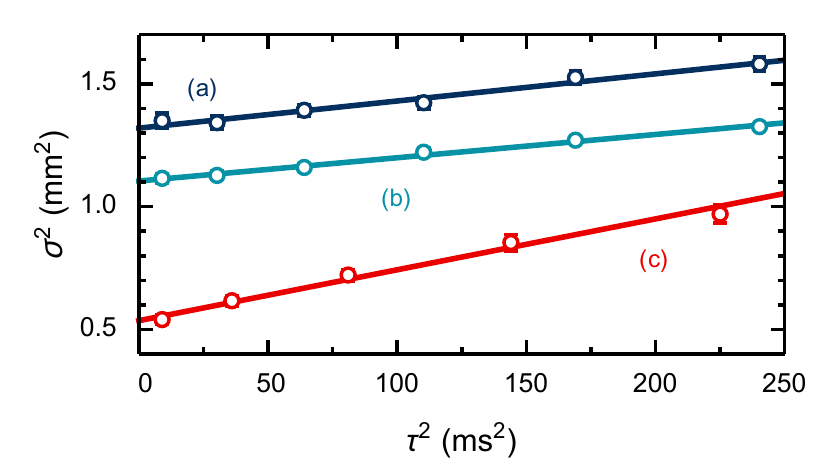}
\caption{Temperature measurements by ballistic expansion. $\sigma$ is rms width of cloud, $\tau$ is expansion time. (a) Axial and (b) radial measurements, after cooling in single-frequency molasses, but prior to compression. The axial and radial temperatures are \SI{8+-1}{\micro\kelvin} and \SI{6.7+-0.6}{\micro\kelvin}. (c) Axial measurement after compression and re-cooling. The compressed rms width is $\sigma_0 = \SI{0.73+-0.01}{\milli\meter}$ and the temperature is $T= \SI{14+-1}{\micro\kelvin}$,  Points: average and standard error of 20 experimental runs. Lines: fits to $\sigma^2 = \sigma_0^2 + k_{\rm B} T \tau^2/m$.}
\label{fig:ballistic}
\end{figure} 

\begin{figure}
\includegraphics{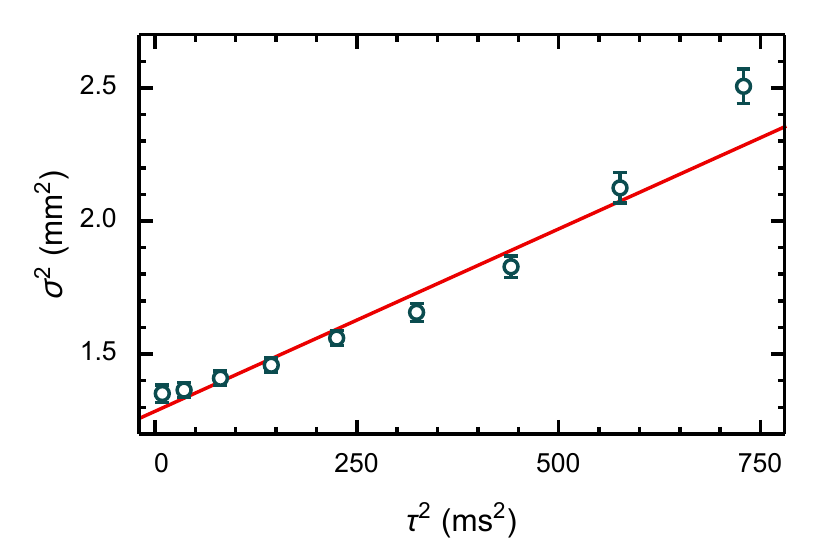}
\caption{Temperature measurement by ballistic expansion, highlighting a difficulty of this method for large, cold clouds. $\sigma$ is rms width of cloud in the axial direction, $\tau$ is expansion time. Points: average and standard error of 20 experimental runs. Line: fit to the model $\sigma^2 = \sigma_0^2 + k_{\rm B} T \tau^2/m$. The data show a systematic deviation from this model, with a lower gradient at earlier times than at late times.\label{fig:ballistic-expansion-systematic}}
\end{figure}

Figure \ref{fig:ballistic}(a,b) shows ballistic expansion measurements of the axial and radial temperatures after applying the single-frequency molasses, but before compressing the cloud. We see that the distribution is cold ($T_{\rm axial}=\SI{8+-1}{\micro\kelvin}$, $T_{\rm radial}=\SI{6.7+-0.6}{\micro\kelvin}$) but large ($\sigma_0 > \SI{1}{\milli\meter}$). As mentioned in the main text,  we find that in this case, the ballistic expansion method of measuring the temperature suffers from systematic errors which are difficult to control. The fundamental problem is that the cloud drops out of the field of view before it has expanded sufficiently for the velocity distribution to be clearly revealed. Figure~\ref{fig:ballistic-expansion-systematic} shows a ballistic expansion dataset extended to longer expansion times, which clearly highlights the problem. At short times, the distribution is dominated by the initial spatial distribution, and deviations from a Gaussian shape (which are common) can lead to systematic shifts in the measurement of the width that change as the cloud expands, potentially altering the gradient of the $\sigma^2$ vs $\tau^2$ plot. Allowing the cloud to expand further reduces this problem, but brings the cloud towards the edges of the field of view as it drops under gravity. Here, imaging aberrations are worse, introducing new systematic uncertainties in the measurement of the cloud size. These effects show up most strongly in the axial expansion data, where we tend to find that the gradient is systematically higher at late times than at early times, as can be seen in Fig.~\ref{fig:ballistic-expansion-systematic}. For this dataset, the temperature determined by fitting to the first four points is \SI{6+-2}{\micro\kelvin}, whilst that found by fitting to the last four points is \SI{14+-1}{\micro\kelvin}. Because there are two potential sources of systematic error that are not easy to separate, and because they depend on the exact shape of the cloud, which is not easily controlled, it is difficult to estimate the size of the systematic correction needed, and the uncertainty on this correction.

Under these conditions, measuring the temperature by the phase-space rotation method developed here is much more precise. Nevertheless, there are still systematic errors to consider. The first comes from the residual difference in the magnetic moments of the molecular states at the offset field used, $\sim \SI{80}{\gauss}$.  As shown in Fig.~\ref{fig:phase-space-rotation}(b)), for this $B_0$ the $(F=2,m_F=-1)$ state has $\mu\approx 0.80~\mu_B$, while the remaining weak-field seeking states of $F=2$ all have $\mu>0.94~\mu_B$. The spread in magnetic moments causes a spread in trap oscillation frequencies which can inflate the minimum size of the cloud, making it appear hotter than it really is. We simulate the effect of this by fitting a single Gaussian to distributions made up of the sum of two Gaussians, one representing the distribution of molecules with $\mu=\mu_B$ in the harmonic trap, and the other, the fraction of molecules with $\mu=0.8~\mu_B$. The effect on the fitted size of the cloud at $t_{1/4}$ depends on the fraction of molecules that have the smaller magnetic moment. Our simulations predict that the molecules are equally distributed amongst the 5 $m_F$ states of $F=2$. This is supported by our observations that 20\% of the molecules are ejected from the magnetic trap (see images in Fig.~\ref{fig:simulations}(d)). Using this distribution, we find a temperature correction of \SI{-0.7+-0.5}{\micro\kelvin} for the coldest clouds. The uncertainty on this correction has been assigned by considering the two extreme scenarios: (i) all molecules having the same magnetic moment, (ii) half the molecules in states of each magnetic moment. This is clearly not a $1\sigma$ bound, but to be conservative we treat it as such when combining with other sources of uncertainty in our analysis. This systematic correction and uncertainty can be reduced by increasing $B_0$.

The finite resolution of our imaging system can make the cloud appear larger than it is. We reduce the severity of aberrations and depth of field effects by aperturing down our imaging system to \SI{20}{\milli\meter} diameter and using only the central $\SI{2.21}{\milli\meter}$ region when integrating over the radial dimension. We have measured the resolution of the imaging optics by imaging a wire of diameter \SI{73}{\micro\meter}. The effect on the images is to convolve the true image with a Gaussian blur of $1/e^2$ radius \SI{0.153+-0.002}{\milli\meter}. This makes all clouds look hotter by \SI{0.67+-0.02}{\micro\kelvin}. Our optics were designed to collect the most light, not to be a good imaging system, and this systematic correction could be significantly reduced with an improved imaging setup. 

The magnification of the imaging system can introduce a systematic error if it is incorrectly calibrated. We determine the magnification to be \num{0.49+-0.01} by fitting the acceleration of a falling cloud to the known acceleration due to gravity. This introduces an uncertainty of \SI{0.2}{\micro\kelvin} for the coldest clouds.

Small errors in determination of the trap frequency have a negligible effect on the measured size of the cloud due to the flatness of $\sigma(t)$ in the vicinity of the minimum. However, an error in the trap frequency results in an error in conversion of the minimum size of the cloud to a temperature. From the oscillation fit in Fig.~\ref{fig:phase-space-rotation}(d) we find this error to be less than \SI{0.1}{\micro\kelvin} for the coldest clouds. 

It is clear from figure~\ref{fig:phase-space-rotation}(d) that the simple model given by Eq.~(\ref{eqn:width}) does not fit perfectly to the oscillation data at late times. The model will be incorrect if the trap is not harmonic. We find that the potential is harmonic to within 0.2\% for $|z|<5$~mm. The maximum rms width of the cloud is only 1.5~mm, so trap anharmonicity has a negligible effect. The trap oscillation frequency, $\omega$, has a small dependence on the radial displacement, $r$. For all $r$ within the selected central region of $w_{\rm cut} = 4.43$~mm, $\omega$ varies by less than 1\%, and this variation has a negligible effect on the fitted model. We simulate the motion of a distribution of molecules using our calculation of the magnetic field due to the coils and the Zeeman shifts of the states, and analyze the change in the spatial distribution over time in the same way as the real data. These simulations show that the method accurately determines the temperature. Using these simulations, we have investigated the potential effect of coupling between the axial and radial motions, and find this has a negligible effect on the temperature measurement. We find that the small difference in magnetic moments of molecules in different states results in slightly different oscillation frequencies which then inflates the minimum size of the cloud, and this inflation increases at later times. Our modelling shows this to be a small, but noticeable, effect that makes a small contribution to the imperfect fit at late times. A larger effect is the defocussing of the cloud as it expands radially out of the imaging plane. We model this effect using data obtained on the resolution of the imaging optics as a function of object position (see above). We find that this defocussing effect is the main cause of the imperfect fit at late times. Because the data points at late times have so little weight in the fit, we did not find it useful to fit a more sophisticated model that includes these effects. 

The red-detuned light used to image the molecules quickly heats the cloud resulting in expansion during the finite exposure time. The effect can be characterised by the size, $\sigma_H$, reached by an initial point source distribution of molecules that is heated for the same amount of time. We find that imaging with blue-detuned light causes negligible expansion for short exposure times at the expense of lower fluorescence and utilise this to determine $\sigma_H$ for different exposure times. Imaging at a detuning of $+3\Gamma/4$ for \SI{750}{\micro\second}, as done throughout this work, gives $\sigma_H=\SI{0.06+-0.08}{\milli\metre}$. As a result, any cloud appears hotter by \SI{0.1+-0.3}{\micro\kelvin}.

Integration of the images over the radial dimension to infer the size in the axial direction introduces the possibility of an error from misalignment of the camera axis relative to the axis of maximum compression. By fitting a two-dimensional Gaussian model with the angle between these axes as a free parameter, we determine this misalignment to be less than \SI{40}{\milli\radian}. This has a negligible effect on the measured sizes of our clouds when using only the central $\SI{2.21}{\milli\meter}$ region of each image in the radial direction, as we do in this work. The effect will be larger at lower temperatures or higher trap frequencies when the cloud is smaller.

Taken together, these systematic shifts in temperature amount to $\SI{-1.4(7)}{\micro\kelvin}$ for the coldest clouds. All the data presented in  this paper have been corrected for the systematic shifts discussed here.

\end{document}